\documentclass[a4paper,fleqn,usenatbib]{mnras}
\usepackage{pdflscape}	% Landscape pages
\usepackage[T1]{fontenc}
\usepackage{ae,aecompl}
\usepackage{amsmath}	% Advanced maths commands
\usepackage{amssymb}	% Extra maths symbols
\usepackage{graphicx}
\begin{document}

\title[The star formation environment of V582 Aur]{The star formation environment of the FU Ori type star 
V582 Aur}

\author[M. Kun et al.]{M. Kun,$^{1}$\thanks{E-mail: kun@konkoly.hu} E. Szegedi-Elek$^{1}$ and 
 B. Reipurth$^{2}$ \\
$^{1}$Konkoly Observatory, Research Centre for Astronomy and Earth Sciences, 
Hungarian Academy of Sciences,\\ 
H-1121 Budapest, Konkoly Thege \'ut 15--17, Hungary \\
$^{2}$Institute for Astronomy, University of Hawaii at Manoa, 640 N. Aohoku Place, Hilo, HI 96720, USA\\}
%\date{Accepted XXX. Received YYY; in original form ZZZ}
\pubyear{2017}
\label{firstpage}
\pagerange{\pageref{firstpage}--\pageref{lastpage}}
\maketitle

\begin{abstract} We have studied the environment of the FU~Ori type star
  V582~Aur. Our aim is to explore the star-forming region associated
  with this young eruptive star. Using slitless spectroscopy we
  searched for H$\alpha$ emission stars within a field of $11.5\arcmin
  \times 11.5\arcmin$, centred on V582~Aur. Based on UKIDSS and
  Spitzer Space Telescope data we further selected infrared-excess young
  stellar object candidates. In all, we identified 68 candidate low-mass young
  stars, 16 of which exhibited H$\alpha$ emission in the slitless 
  spectroscopic images. The colour--magnitude diagram of
  the selected objects, based on IPHAS data, suggests that they are
  low-mass pre-main-sequence stars associated with the Aur~OB\,1
  association, located at a distance of 1.3~kpc from the Sun. The 
  bright-rimmed globules in the local environment of V582~Aur 
  probably belong to the dark cloud LDN~1516. Our results suggest 
  that star formation in these globules might have been 
  triggered by the radiation field of a few hot members of Aur~OB\,1.
  The bolometric luminosity of V582~Aur, based on archival photometric
  data and on the adopted distance, is 150--320\,L$\sun$.
\end{abstract}


\section{Introduction}

FUors, named after their prototype FU Orionis, are young, low-mass
stars undergoing powerful, long term (from decades to centuries)
outbursts, powered by increased accretion \citep{hartmann1996,audard}.
During these events, FUors brighten up to 6 magnitudes over a few
months, stay in a high state for decades, and as much as one tenth of
the stellar mass may be added during these repeated outbursts.  These
young stars are members of star-forming regions both
kinematically and spatially
\citep{herbig1966,herbig1977,hartmann1996}. Studying the star forming
environment of FUors is important for determining their distances and
luminosities.

One of the less studied FU~Ori type stars is V582~Aur
\citep{samus2009,semkov2013}. The brightening of V582~Aur was discovered by
the amateur astronomer Anton Khruslov and the star was reported as a FUor
candidate by \citet{samus2009}. \citet{semkov2013}, based on photometric and 
spectroscopic observations of V582~Aur in its high state,
confirmed its FUor nature. V582~Aur is projected on both the
Aur~OB\,1 and Aur~OB\,2 associations, which are overlapping along the line of
sight and situated at 1.32$\pm0.1$ and 2.8$\pm0.27$~kpc from the Sun,
respectively \citep{humphreys1978,marco2016}. The region was included 
in several observational studies of the Galactic structure toward the 
anticentre \citep*[e.g.][and references therein]{marco2016,camargo2013}.
V582~Aur is projected within the area  of the cluster CBB~9, identified
by \citet*{camargo2012}, and on the periphery of the large dark cloud Lynds~1516 \citep{LDN}. 
No molecular cloud associated with Lynds~1516 was listed in \citet{Kawamura98}.

In order to explore the star formation environment of V582~Aur we
searched for candidate young stellar objects in the region centred on
V582~Aur. We utilized two striking characteristics of low- and
intermediate-mass young stars, i. e. their specific near- and
mid-infrared excesses, originating from circumstellar discs, and
strong H$\alpha$ emission, a signpost of mass accretion. We performed
a search for new H$\alpha$ emission line stars via slitless grism
spectroscopy, using the University of Hawaii 2.2-m telescope, and used
infrared data, available in the UKIDSS Galactic Plane Survey data base
\citep{Lucas2008} and in the \textit{Spitzer\/} GLIMPSE360 Catalog and
Archive. We further used optical photometric data, available in the IPHAS
\citep{Barentsen2014} DR2 Source Catalogue, to estimate distances and
luminosities. We describe our observations and data reduction in
Sect.~\ref{Sect_2}. The procedures for young stellar object (YSO)
identification in archival data are described in Sect.~\ref{Sect_3}.
The results are presented and conclusions are drawn in
Sect.~\ref{Sect_4}. We give a short summary in Sect.~\ref{Sect_5}.

\section{Observations and data reduction}
\label{Sect_2}

We conducted a search for H$\alpha$ emission stars in the vicinity
of V582~Aur using the Wide Field Grism Spectrograph 2 (WFGS2) at the
University of Hawaii 2.2-meter telescope on 2011 January 1. We used a
300 line mm$\sp{-1}$ grism blazed at 6500~\AA\ and providing a
dispersion of 3.8~\AA\,pixel$\sp{-1}$ and a resolving power of 820.
The H$\alpha$ filter had a 500~\AA\ passband, centred near 6515~\AA.
The detector for WFGS2 was a Tektronix 2048$\times$2048 CCD, whose
pixel size of 24\,$\mu$m corresponded to 0.34~arcsec on the sky. The
field of view was $11.5\arcmin\times11.5\arcmin$, centred on V582~Aur. 
We took a short, 60\,s exposure to identify the H$\alpha$ line in the 
spectra of bright stars, and avoid saturation. Then we obtained two frames 
of 480\,s exposure time, and co-added them to eliminate effects of the 
cosmic rays. Direct images of the same field were obtained through
$r^{\prime}$ and $i^{\prime}$ filters before the spectroscopic
exposures. One exposure was taken in each filter with integration
time of 60\,s. The steps of the data reduction were the same as
described in detail in \citet{szegedi}.

We examined the co-added spectroscopic image visually to discover 
young stars showing the H$\alpha$ line in emission, and found 
16 emission-line objects in the observed region. 
We determined their equatorial coordinates by matching their
positions, measured in the coadded grism image, with those of 
Two Micron All Sky Survey \citep{cutri2003} point sources in the direct images. 
All H$\alpha$ emission sources had
2MASS counterparts within 1.4 arcseconds. We use 2MASS designations to
identify the new H$\alpha$ emission objects. The equivalent width of
the H$\alpha$ emission line EW(H$\alpha$) and its uncertainty were
computed in the manner described by \citet{szegedi}. Due to the faint
continuum and/or overlapping spectra we could measure EW(H$\alpha$)
only in the spectra of nine stars. Table~\ref{table1} lists the
2MASS designations and derived H$\alpha$ equivalent widths of the
emission objects detected in the WFGS2 images centred on V582~Aur. For
comparison, we list EWs estimated from the IPHAS $r^\prime-[$H$\alpha]$ vs.
$r^\prime-i^\prime$ colour--colour diagram (see Sect.~\ref{Sect_pms}) 
in column~5. To indicate the brightness of the stars $r^\prime$ 
magnitudes are listed in column~6 (see Sect.~\ref{Sect_iphas}). The last 
column of Table~\ref{table1} indicates the presence or 
absence of infrared excess over the 2.2--4.6\,$\mu$m wavelength region
(see Sects.~\ref{Sect_ukidss} and \ref{Sect_spitzer}). 

\begin{table*}
\caption{H$\alpha$ emission stars identified during our \textit{WFGS2} observations.}
\label{table1}
\begin{center}
\begin{tabular}{lccccc}
\hline\hline
\noalign{\smallskip}
Name & 2MASS Id  &  \multicolumn{2}{c}{EW(H$\alpha$) (\AA)} & $r^\prime_\mathrm{IPHAS}$ & IR excess \\ 
%  & & \multicolumn{2}{c}{(\AA)} & (mag) \\ 
  &&  WFGS2$^{a}$ & IPHAS$^{b}$ & (mag)\\
\hline
 L1516-Ha\,1 & 05252786+3457419  &  $\cdots$      & ~25    & 19.02 & Y \\    
 L1516-Ha\,2 & 05252963+3457551  &  8.0$\pm$2.5   & ~25   & 17.78 & Y \\ 
 L1516-Ha\,3 & 05253251+3457180  &  11.6$\pm$2.3  & ~15    & 17.16 & Y \\
 L1516-Ha\,4 & 05253272+3457107  &  13.7$\pm$2.2   & ~50    & 16.51 & Y  \\
 L1516-Ha\,5 & 05253284+3455293  &  $\cdots$	  & $<10$~  & 18.52 & N \\
 L1516-Ha\,6 & 05253408+3457465  &  $\cdots$	  & ~60    & 19.09 & Y \\
 L1516-Ha\,7$^{*}$ & 05254456+3450178  &  53.3$\pm$10.5 & ~30  & 17.85 & Y \\
 L1516-Ha\,8 & 05255203+3458081  &   6.8$\pm$0.7  & ~80    & 18.36 & Y \\
 L1516-Ha\,9$^{*}$ & 05255439+3454391  &  37.3$\pm$4.2  & ~50 & 17.45 & Y \\
L1516-Ha\,10 & 05255537+3454418  &   6.5$\pm$2.1  & $<10$~   & 17.96 & Y \\
L1516-Ha\,11 & 05255688+3451442  &  46.6$\pm$5.6  & 100   & 16.96 & Y \\
L1516-Ha\,12 & 05260273+3451093  &  $\cdots$	  & 160   & 20.41 & Y \\
L1516-Ha\,13 & 05260711+3454260  &  $\cdots$	  & $<10$~  & 18.00 & Y \\
L1516-Ha\,14 & 05261468+3450052  &  $\cdots$	  & ~60    & 19.73 & Y \\
L1516-Ha\,15 & 05261637+3453332  &  15.4$\pm$7.7  & $<10$~  & 17.72 & Y \\
L1516-Ha\,16 & 05261681+3454223  &  $\cdots$	  & 120   & 20.99 & Y \\
\hline
\end{tabular}
\end{center}
\smallskip
\flushleft{\small 
$^{a}$Measured in the slitless grism spectra. 
$^{b}$Estimated from IPHAS photometry (Fig.~\ref{fig_iphas1}).
$^{*}$Catalogued as H$\alpha$ emission star in \citet{witham2008} from IPHAS 
photometry.}
\end{table*}

\section{YSO selection from data archives}
\label{Sect_3}
\subsection{UKIDSS}
\label{Sect_ukidss}
Pre-main-sequence stars, surrounded by protoplanetary discs, are
clustered at specific regions in the {\it J\/}$-${\it H\/} vs. {\it H\/}$-${\it K}$_{\mathrm{s}}$
colour--colour diagram. Thus the colour--colour diagram is a useful tool to
identify young stars. The region around V582~Aur was covered by the
UKIDSS Galactic Plane Survey, some three magnitudes deeper than the
2MASS survey. We used the UKIDSS-DR6 Galactic Plane Survey data base
\citep[UGPS,][]{Lucas2008}, accessible in
\textit{VizieR\/}\footnote{http://vizier.u-strasbg.fr/}, to identify 
disc-bearing young stars in the area, identical with the field of view 
of our WFGS2 observations.  Figure~\ref{fig_select1} presents the colour--colour 
diagram of the UKIDSS sources of the studied region. 
Small grey crosses show the distribution of all UKIDSS 
sources detected in each band, classified as stars with 
a probability pStar > 0.70, and having colour index uncertainties 
$\Delta$({\it J}$-${\it H\/})$<0.1$ and $\Delta(${\it H}$-${\it K}$_\mathrm{s})<0.1$.
Candidate YSOs in this diagram are sources whose 
errorbars are entirely to the right of the band of the reddened main sequence, 
bordered by the long-dashed lines. Reddened T~Tauri stars occupy 
the grey area \citep*{meyer1997}. Taking into account the empirical nature of 
the T~Tauri locus (dash-dotted line), we include a 0.1-mag wide band below 
this line into the T~Tauri domain. Sources to the right of this
band are candidate embedded protostars, whereas those below the locus of 
the unreddened T~Tauri stars are somewhat uncertain in nature. 
Herbig~Ae/Be stars are located in this part of the diagram
\citep[e.g.][]{manoj2006}. These intermediate-mass pre-main-sequence stars
are expected to be brighter than T~Tauri stars of the same star-forming
region. We checked the magnitudes of the {\it K}$_\mathrm{s}$-band-excess 
sources below the unreddened T~Tauri locus and found them to be among the
faintest sources of the studied field. We ignore these sources, since
their very red {\it H\/}$-${\it K}$_\mathrm{s}$ colour indices probably do not 
originate from protoplanetary discs. Red diamonds indicate the 42 selected 
candidate YSOs in Fig.~\ref{fig_select1}. Additionally, twenty extended UKIDSS 
sources of the studied area, classified as probable galaxies, are located in 
the same region of the colour--colour plane. Extended sources, projected on a 
star-forming region, may be embedded protostars, whose appearance in the 
near-infrared is dominated by scattered light. We examined these sources 
in the UKIDSS and IPHAS images, and added six of them to the list of 
candidate YSOs. Black circles indicate these six extended sources in 
Figs.~\ref{fig_select1} and \ref{fig_map1}. The list of the UKIDSS-selected 
candidate YSOs is presented in Table~\ref{table2}.
Ten H$\alpha$ stars, detected in the WFGS2 images, have counterparts
in this sample. V582~Aur itself is saturated in the UKIDSS images. Its
2MASS data, however, show it to be a {\it K}$_\mathrm{s}$-band excess young star.

\subsection{{\it Spitzer}}
\label{Sect_spitzer}
Our target field was also observed by the \textit{Spitzer Space Telescope\/}
\citep{werner2004} during its warm mission, as part of the GLIMPSE360
project \citep{churchwell2009}. Observations were performed on 2010
May 4 with the Infrared Array Camera \citep[IRAC,][]{fazio2004} at 3.6
and 4.5\,$\mu$m (aors 38916864 and 38828288). We examined all sources
within the $11.5\arcmin\times11.5\arcmin$ environment of V582~Aur to
search for additional young stars. We combined UKIDSS
$JHK_\mathrm{s}$ magnitudes with \textit{Spitzer\/} [3.6] and [4.5] magnitudes
and applied the \textit{Phase~2} selection criteria established by
\citet{gutermuth2009}. This selection process includes dereddening of
the colour indices of input stars onto the YSO locus of the $J-H$ vs.
$H-K_\mathrm{s}$ (or {\it H\/}$-${\it K}$_\mathrm{s}$ vs. $[4.5]-[3.6]$ when {\it J\/}-band 
photometry is not available) diagram, and defines YSO colour criteria 
in the plane of the extinction-corrected $[K_\mathrm{s}-[3.6]]_0$ and
$[[3.6]-[4.5]]_0$ colour indices. Fig.~\ref{fig_select2} shows the
distribution of the point sources in the $[K_\mathrm{s}-[3.6]]_0$ vs.
$[[3.6]-[4.5]]_0$ colour--colour diagram. To exclude dim
extragalactic contaminants we applied $[3.6] < 14.0$ and $[4.5] < 14.0$
magnitude limits. Candidate YSOs are located in the upper right part
of the diagram, bordered by the solid lines. Blue squares indicate the
48 candidate YSOs revealed by this selection. For comparison we plotted
with red dots the candidate YSOs selected in the UKIDSS {\it J\/}$-${\it H\/} vs.
{\it H\/}$-${\it K}$_\mathrm{s}$ colour--colour diagram. The comparison shows that
each previously selected object lies within the region assigned for
YSOs, and a sizable part of them are fainter than 14~mag in the IRAC
bands.  We find 16 candidate YSOs not selected in the UKIDSS data. 
These stars are listed in Table~\ref{table3}. Their positions in the 
{\it JHK}$_\mathrm{s}$ colour--colour diagram are indicated by blue squares 
in Fig.~\ref{fig_select1}. A 17th source, G172.7687-00.3577
(UGPS~J052547.76+344950.9), classified as a probable UKIDSS galaxy, also
fulfils the colour criteria. This very red object is projected on 
the edge of a dark cloud, therefore we include it in the list of 
candidate YSOs. Five H$\alpha$ stars, detected in the WFGS2 images, have 
counterparts in this sample. 

\subsection{{\it WISE}}\label{Sect_wise}

\citet{Marton2016} performed a comprehensive all-sky search for
candidate YSOs in the \textit{AllWISE\/} data release. Six objects of
their resulting catalogue \citep{Marton2016a} can be found in our
studied area, including V582~Aur itself. Three further among these
candidates coincide with sources selected during the previous steps.
These stars are marked in Tables~\ref{table2} and \ref{table3}. 
  The remaining two objects are new candidate YSOs, listed in
  \citeauthor{Marton2016}'s \citeyearpar{Marton2016} catalogue as
  AllWISE~J052614.87+345225.0 and AllWISE~J052538.15+344805.0. The
  first source was classified as a probable galaxy (extended source)
  in the UKIDSS and IPHAS bands, therefore its nature is uncertain.
  The UKIDSS and \textit{Spitzer\/} sources, coinciding in position
  with the second source within one arcsec, exhibit normal stellar
  colours, indicative of either an evolved (transitional)
  circumstellar disc, or two distinct sources.

\subsection{IPHAS} \label{Sect_iphas}
The field we have studied was covered by the INT Photometric H$\alpha$ Survey
of the Northern Galactic Plane (IPHAS) survey \citep{Drew2005}.
High-quality (S/N > 10) $r^\prime$ and $i^\prime$, and narrow-band
H$\alpha$ magnitudes are available for more than 800 stars of the
studied $11.5\arcmin\times11.5\arcmin$ region in the IPHAS DR2 Source Catalogue
\citep{Barentsen2014}. These data are suitable for selecting further
candidate YSOs \citep[see][]{Barentsen2011}. Two of our 16 H$\alpha$
emission stars appear in the first catalogue of H$\alpha$ emission
objects based on the IPHAS survey \citep{witham2008}, containing
stars brighter than $r^\prime = 19.5$~mag. $r^\prime$, $i^\prime$, and
H$\alpha$ magnitudes are available for 35 of our candidate YSOs. We
use these data to examine the H$\alpha$ emission properties of the
stars selected by infrared excesses, and establish the
distance of the group of candidate YSOs around V582~Aur.

\begin{figure}
\resizebox{\hsize}{!}{\includegraphics{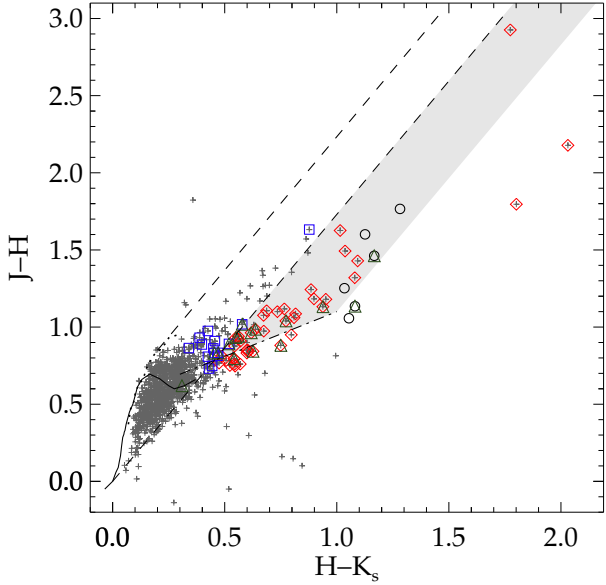}}
\caption{{\it J\/}$-${\it H\/} vs. {\it H\/}$-${\it K}$_\mathrm{s}$ colour--colour 
diagram of UKIDSS sources in a field of $11.5\arcmin\times11.5\arcmin$, 
centred on V582~Aur. Small grey crosses indicate all stars having 
$\Delta$({\it J\/}$-${\it H\/})$<0.1$ and $\Delta$({\it H\/}$-${\it K}$_\mathrm{s}$)$ <0.1$ in the UGPS catalogue. 
Red diamonds mark the candidate YSOs selected out of this sample.
Green triangles show the H$\alpha$ emission stars identified in the WFGS2 
images. Black circles show the extended UKIDSS sources considered as
candidate YSOs. The black solid curve shows the colours of the zero-age 
main sequence, and dotted line is the giant branch. The dash-dotted line 
is the locus of unreddened T~Tauri stars \citep{meyer1997}. 
The long-dashed lines delimit the area occupied by reddened normal 
stellar photospheric colours \citep*{cardelli1989}. The grey shaded band 
indicates the area occupied by reddened {\it K\/}$_\mathrm{s}$-band 
excess pre-main-sequence stars. For comparison, we plotted with 
blue squares candidate disc-bearing stars revealed by their 3.6 
and 4.5-$\mu$m excess emission only.}
\label{fig_select1}
\end{figure}

\begin{figure}
\resizebox{\hsize}{!}{\includegraphics{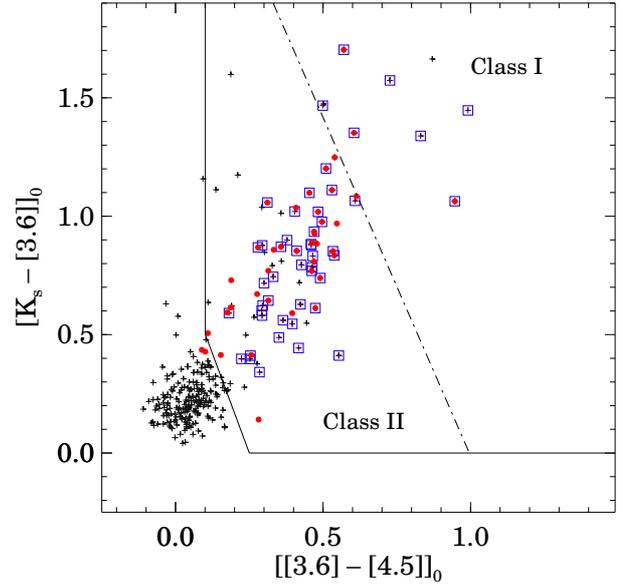}}
\caption{[{\it K}$_\mathrm{s}-[3.6]]_{0}$ vs. [$[3.6]-[4.5]_{0}]$  colour--colour diagram of 
all \textit{Spitzer\/} sources associated with UKIDSS sources and brighter 
than $[3.6]=14.0$~mag and $[4.5]=14.0$~mag in the $11.5\arcmin\times11.5\arcmin$ 
area centred on V582~Aur. Solid lines delimit the region occupied by 
candidate YSOs, and the dash-dotted line separates Class~I protostars from 
Class~II pre-main-sequence stars according to \citeauthor{gutermuth2009}'s~\citeyearpar{gutermuth2009} 
\textit{Phase~2\/} method. Blue squares indicate candidate YSOs identified 
in this diagram, and red dots correspond to the sources selected by their 
{\it K}$_\mathrm{s}$-band excess.}
\label{fig_select2}
\end{figure}

\section{Results}
\label{Sect_4}

\subsection{Surface distribution of candidate YSOs around V582~Aur}
\label{Sect_map}
\begin{figure}
%\sidecaption
\resizebox{\hsize}{!}{\includegraphics{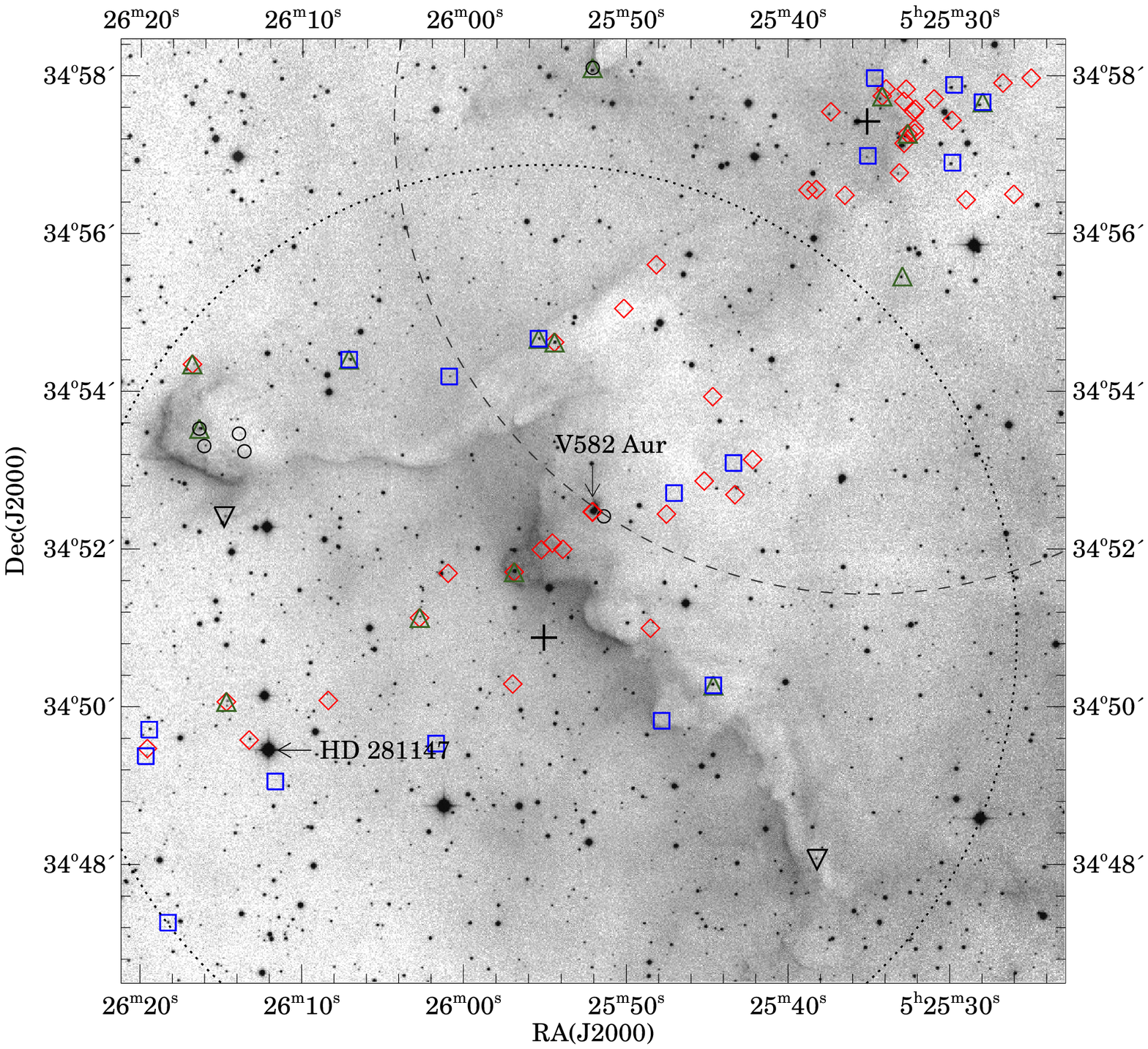}}
\caption{Candidate young stars identified during the present survey, overplotted
on the IPHAS narrow-band H$\alpha$ image. Green triangles represent H$\alpha$ 
emission stars, red diamonds mark UKIDSS-selected sources, and blue squares 
indicate candidate pre-main-sequence stars exhibiting excess emission only 
in the \textit{Spitzer\/} data. Small black circles indicate extended 
UKIDSS sources considered as candidate YSOs. Inverted triangles show the
\textit{AllWISE\/} sources listed in \citeauthor{Marton2016}'s 
\citeyearpar{Marton2016} catalogue of candidate YSOs. The large dotted 
circle indicates the position of the CBB\,9 cluster, and the dashed circle 
marks the position and radius of the FSR\,775 open cluster candidate. 
Crosses show the catalogued positions of the cluster centres.}
\label{fig_map1}
\end{figure}

Figure~\ref{fig_map1} shows the surface distribution of the selected
candidate young stars, overplotted on the IPHAS narrow-band H$\alpha$
image of the studied $11.5\arcmin\times11.5\arcmin$ region. The underlying 
H$\alpha$ image reveals that V582~Aur and several candidate young stars 
are projected on bright-rimmed dark clouds. Striking features of the image 
are the elephant-trunk-like globule near the eastern edge, and the compact
group of candidate YSOs clustered at the north-western corner. 
The dotted circle indicates the catalogued position of the CBB\,9 cluster, 
identified by \citet{camargo2012}, and the dashed circle marks the position 
of the FSR\,775 open cluster candidate, identified by
\citet*{froebrich2007} using 2MASS data, and rejected later as a real cluster 
by detailed structure studies by \citet{camargo2012}. The distribution of fainter 
UKIDSS and \textit{Spitzer\/} sources suggest a remarkable clustering of 
candidate T~Tauri stars projected near the centre of FSR\,775. Some 40~per cent 
of the candidate YSOs can be found within 2~arcmin to the centre of FSR~775. 
Study of the radial profile of this group, however, is beyond the scope of 
the present paper. Our candidate young stars and V582~Aur itself are 
projected within these overlapping clusters. The B1\,V type star 
HD~281147 \citep{marco2016}, mentioned by \citet{camargo2012} as a possible 
member of CBB\,9 is also labelled.

\subsection{The nature of candidate YSOs associated with V582~Aur}
\label{Sect_pms}
All but one of the H$\alpha$ emission stars detected by the WFGS2 exhibit 
infrared excesses characteristic of classical T~Tauri stars. L1615-Ha\,5
has no excess flux in the wavelength region covered by the UGPS and 
\textit{Spitzer\/} GLIMPSE observations. Its low-quality \textit{AllWISE\/} fluxes 
do not exclude excess emission at longer wavelengths. Its nature is thus 
uncertain: it may be either a pre-main-sequence star with evolved 
(transitional) disc, or weak-line T~Tauri or M-type main sequence star 
with H$\alpha$ emission. Another H$\alpha$ source of uncertain nature is 
L1516-Ha~1, whose associated UKIDSS source was classified as a galaxy. 
 
\begin{figure}
\resizebox{\hsize}{!}{\includegraphics{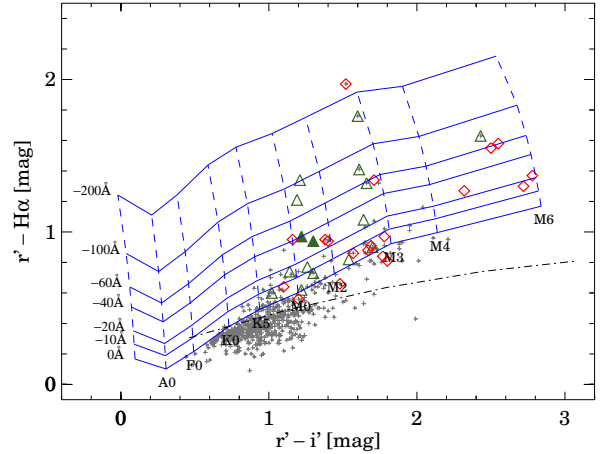}}
\caption{$r^\prime-[$H$\alpha]$ vs. $r^\prime-i^\prime$ colour--colour diagram of 
35 candidate YSOs within the $11.5\arcmin\times11.5\arcmin$  box around 
V582~Aur. Triangles indicate H$\alpha$ emission stars identified in our 
WFGS2 images, and diamonds mark infrared-excess stars. Filled triangles 
mark the stars included in the IPHAS emission star catalogue 
\citep{witham2008}. Small grey crosses show the colour distribution of 
the IPHAS sources with $S/N > 10$ in each band within the same area. 
Simulated colours of main sequence stars with H$\alpha$ emission, 
computed by \citet{Barentsen2011} are indicated by the solid blue lines 
for $E(B-V)=0.40$, the lower limit of foreground reddening towards the line 
of sight of V582~Aur. Dashed blue lines show the colour variations of the 
spectral types indicated at the lower ends, due to increasing EW(H$\alpha$).
Dash-dotted line shows the reddening path of a K0 type star between 
$0 \leq E(B-V)\leq 4$. }
\label{fig_iphas1}
\end{figure}

The $r^\prime-[$H$\alpha]$ vs. $r^\prime-i^\prime$ colour--colour 
diagram allows us to estimate the EW of the H$\alpha$ emission line of 
the studied stars \citep{Barentsen2011}.
Figure~\ref{fig_iphas1} shows this diagram for the 35 candidate YSOs
detected in each IPHAS band. Synthetic colours of main sequence stars and
H$\alpha$ emission objects are taken from \citeauthor{Barentsen2011}'s
\citeyearpar{Barentsen2011} Table~A1. The thick solid line indicates
the normal main sequence, and thin solid lines show the main sequence
colours modified by H$\alpha$ emission. Each colour was reddened to
E(B$-$V) = 0.40~mag, the lower limit of the foreground reddening towards
the line of sight of V582~Aur \citep{green2015}. It can be seen that not
all of our candidate YSOs appear as H$\alpha$ emission stars in this
diagram. Nearly half of them (15 objects), including five of the 
WFGS2-detected H$\alpha$ emission stars are scattered around the main 
sequence, below the line corresponding to EW(H$\alpha$)=$-10$\,\AA. 
All of these stars, except L1615-Ha\,5, have CTTS-like infrared excesses. 
Their positions may result from their significantly higher 
extinctions, or they may be variable classical T~Tauri stars with 
temporarily low accretion activities. 
Figures~\ref{fig_iphas1} and \ref{fig_iphas2} (Sect.~\ref{Sect_cmd}) suggest 
that the L1516-Ha objects with weak H$\alpha$ emission line are bluer 
and brighter on the average than the whole sample of the candidate YSOs. 
These properties suggest reddened G and early K type T~Tauri stars, in 
which accretion may result in EW(H$\alpha$) below the $-10$\,\AA\  threshold 
\citep{Barrado2003}.

\begin{figure*}
\centerline{\includegraphics{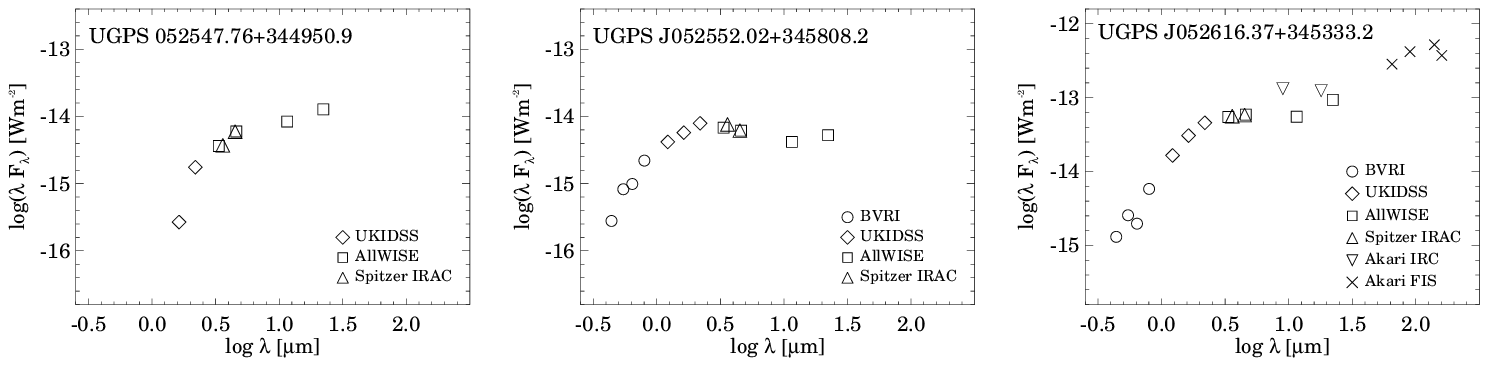}}
\caption{Spectral energy distributions of three candidate protostars in the
environment of V582~Aur. Optical data are from the NOMAD Catalog and IPHAS.} 
\label{fig_sed1}
\end{figure*}

Figures \ref{fig_select1} and \ref{fig_select2} suggest that a few
selected sources may be Class~I protostars \citep{Lada1991}. To confirm their
nature we looked for mid- and far-infrared counterparts in the AllWISE, 
Akari IRC and Akari FIS data. Spectral energy distributions (SED)
for three candidate protostars are displayed in Fig.~\ref{fig_sed1}.
UGPS~J052547.76+344950.9 is a faint, red source not detected in the 
UKIDSS {\it J\/} band. Its 22-$\mu$m flux indicates a Class~I protostar.
UGPS~J052552.02+345808.2 is projected near the bright rim 
of a small cometary globule at the northern edge of the region. It is 
included in the catalogue of WISE YSO Candidates \citep{Marton2016a}. Its SED
suggests a flat-spectrum object. UGPS~J052616.37+345333.2 (L1516-Ha\,15) is 
situated near the centre of the elephant-trunk-like cometary globule.
Its SED suggests a Class~I YSO. Another nearby candidate Class~I YSO, 
UGPS~J052616.07+345320.0, may contribute to the Akari FIS fluxes. 

\subsection{Colour--magnitude diagram: the distance of V582~Aur} 
\label{Sect_cmd}

\begin{figure}
\resizebox{\hsize}{!}{\includegraphics{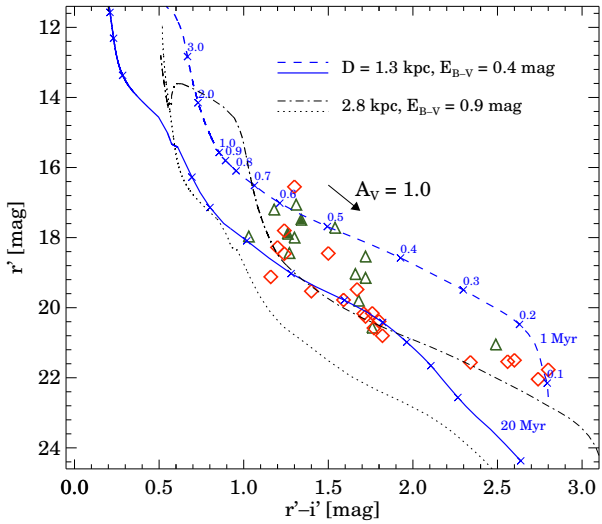}}
\caption{$r^\prime$ vs. $r^\prime-i^\prime$ colour--magnitude diagram of the 35 
candidate YSOs detected by IPHAS. Symbols are same as in Fig.~\ref{fig_iphas1}. 
1 and 20-Myr isochrones \citep{bell2014} are drawn for the distances 
and foreground extinctions of Aur~OB\,1 (blue) and Aur~OB\,2 (black). 
Stellar masses between 0.1 and 3.0\,M$_{\sun}$ are marked along the 
isochrones and labelled along the 1-Myr line.}
\label{fig_iphas2}
\end{figure}

When plotting the $r^\prime$ vs. $r^\prime-i^\prime$ colour--magnitude
diagram one has to keep in mind that $r^\prime$ magnitudes may be
modified by the presence of the H$\alpha$ emission line within the
$r^\prime$ band \citep{Drew2005,Barentsen2011}. We applied the correction to 
the $r^\prime$ magnitudes following \citet{Barentsen2011}, and plotted the 
$r^\prime$ vs. $r^\prime-i^\prime$ colour--magnitude diagram of the 
candidate YSOs in Fig.~\ref{fig_iphas2}. We compare their distribution 
with semi-empirical pre-main-sequence isochrones
presented by \citet{bell2014} for the IPHAS bands, based on the Pisa 
pre-main sequence tracks and isochrones \citep{tognelli2011} and BT-Settl 
\citep[][and references therein]{BCAH15} atmosphere models. 
Isochrones bracketing from $10^{6}$ to $2\times10^{7}$-year encompasses
almost all disc-bearing pre-main-sequence stars, therefore
we plotted in Fig.~\ref{fig_iphas2} isochrones for the distances and foreground 
extinctions of both Aur~OB\,1 and Aur~OB\,2. Foreground reddenings of 
$E(B-V)=0.40$~mag and $E(B-V)=0.90$~mag were adopted for 
Aur~OB\,1 and Aur~OB\,2, respectively \citep{green2015}. 

Most of the candidate YSOs in this diagram are widely scattered between the
isochrones plotted for the distance and foreground extinction of Aur~OB\,1,
and above or along the 1-Myr isochrone of Aur~OB2. Taking individual 
extinctions into consideration would modify this distribution so that 
positions of the stars reddened by $E(B-V) > 0.40$~mag would move 
toward slightly younger isochrones and higher masses. 
The colour--magnitude diagram suggests the conclusion that the star-forming 
region around V582~Aur is probably associated with Aur~OB1, located at 
1.32\,kpc from the Sun \citep{humphreys1978}.

%\section{Discussion}
%\label{Sect_5}

\subsection{Bolometric luminosity of V582~Aur}

\begin{figure}
\resizebox{\hsize}{!}{\includegraphics{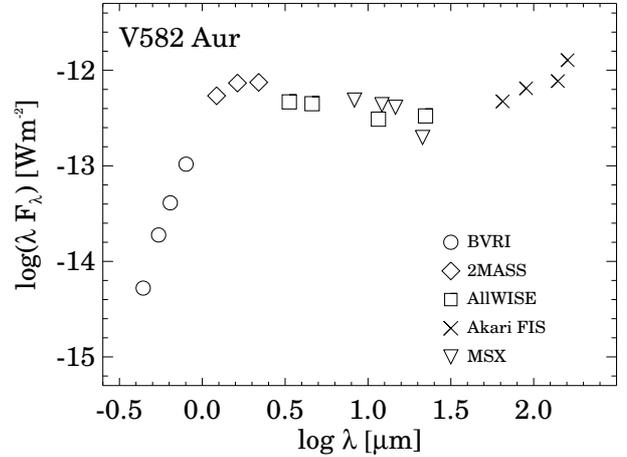}}
\caption{Spectral energy distribution of V582~Aur, based on archival data. 
$BVRI$ data are averaged from \citeauthor{semkov2013}'s \citeyearpar{semkov2013} 
light curves.} 
\label{fig_v582}
\end{figure}

Our results suggest that the FUor V582~Aur is a member of a group of
young low-mass stars, situated at a distance of 1.32~kpc from the Sun.
The adopted distance allows us to estimate the luminosity of this
outbursting young star. The SED of V582~Aur, constructed from archival
data, is displayed in Fig.~\ref{fig_v582}. The far-infrared source
Akari FIS 0525509+345227, associated with V582~Aur in the Akari~FIS
YSO Catalogue of \citet{Toth2014}, is separated by 13.5\arcsec\ from
the star. Although this separation is smaller than the half-maximum
radius of the point-spread function of the FIS \citep{Arimatsu2014},
unresolved nearby sources may contribute to the far-infrared fluxes.
We integrated the SED to obtain the bolometric luminosity of V582~Aur,
after correcting the fluxes for a foreground extinction
$A_\mathrm{V}=1.53$~mag, taken from the IPHAS 3D extinction map of the
Northern Galactic Plane (http://www.iphas.org/extinction/), and using 
\citeauthor{cardelli1989}'s \citeyearpar{cardelli1989} extinction law. We
obtained $L_\mathrm{bol}=324$\,L$_{\sun}$ with the FIS data included,
and $L_\mathrm{bol}=146$\,L$_{\sun}$ without the far-infrared part of
the SED. These values are typical for FUor outburst luminosities
\citep{audard}.

\subsection{Connection of the new star-forming region with Aur~OB\,1}\label{Sect_large}

The colour--magnitude diagram of the candidate YSOs around V582~Aur 
suggests that their associated cometary globules and bright-rimmed dark clouds 
(Fig.~\ref{fig_map1}) are located within the volume of the Aur~OB\,1 associaton.
The bright rims indicate interaction of the clouds with hot stars of
the association. Aur~OB\,1 is defined by a few O and early B type stars 
\citep{humphreys1978}, and the 20-Myr old open cluster
NGC\,1960 \citep{Reipurth2008}. \citet*{straizys2010} established that the dark cloud 
LDN~1525 (TGU 1192), located some 2~degrees north-east from our studied 
area and associated with the \ion{H}{II} region Sh2-235, is situated 
at 1.3~kpc from the Sun, within the volume occupied by Aur~OB\,1. Several 
signposts of on-going star formation were identified in the region of 
Sh2-235 \citep[][and references therein]{Dewangan2016}, which may be a possible 
young subsystem of Aur~OB\,1. The complex of bright-rimmed clouds, associated 
with V582~Aur, may represent another region of active star formation within 
the volume of Aur~OB\,1. 

\begin{figure}
\resizebox{\hsize}{!}{\includegraphics{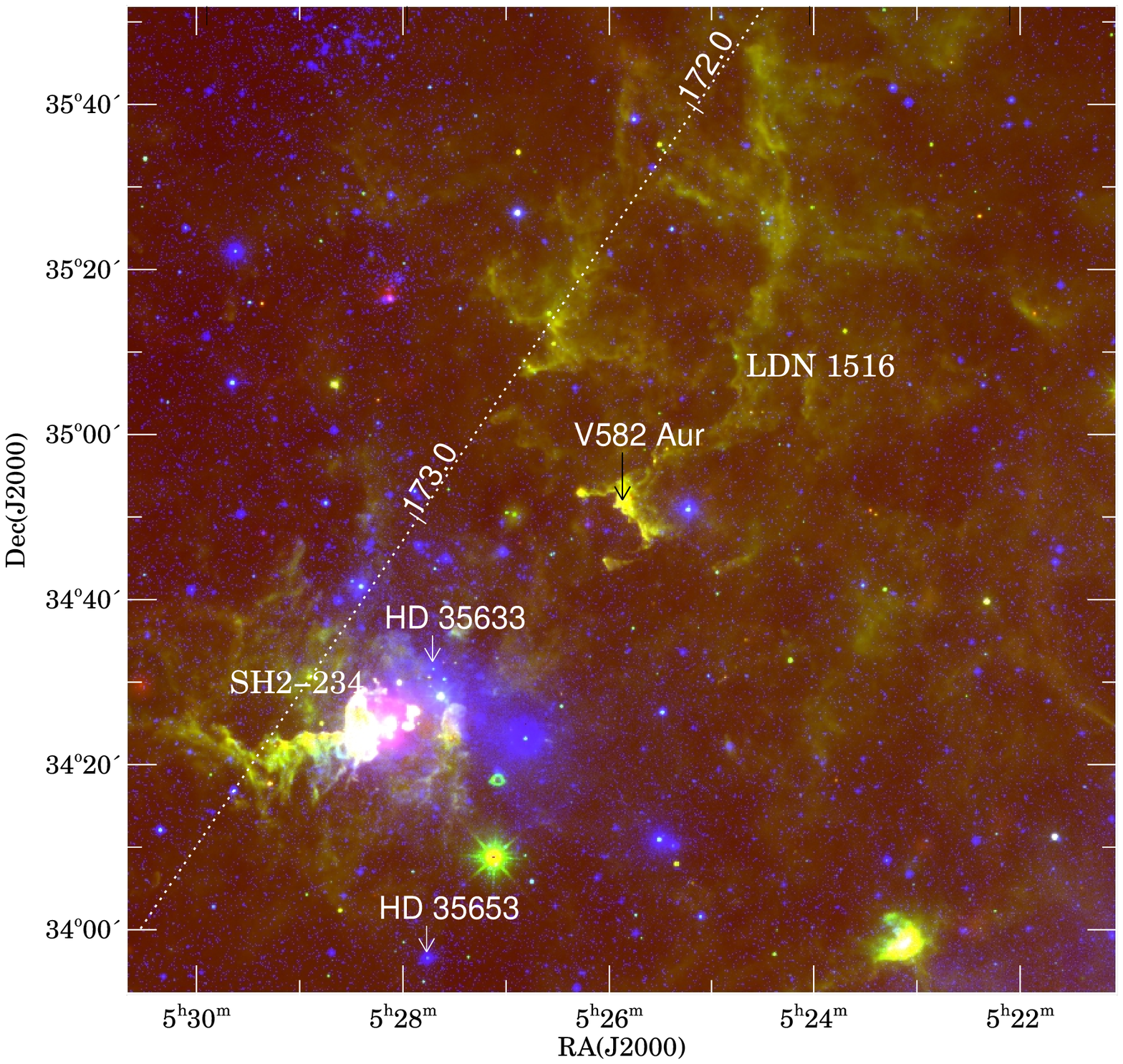}}
%\vskip-3cm
\caption{Three-colour image of the 2$\degr\times2\degr$ environment of V582~Aur,  
composed from the  DSS2b (blue), \textit{WISE\/} 12\,$\mu$m (green), and 
\textit{WISE\/} 22\,$\mu$m (red) images. The newly identified bright-rimmed 
clouds can be seen in the middle of the image. The possible exciting stars 
are labelled. The dotted line shows the Galactic equator.}
\label{fig_map2}
\end{figure}

To find the possible ionizing stars we examined the wide-field
environment of V582~Aur. Figure~\ref{fig_map2} presents a	         
$2\degr\times2\degr$ three-color image, centred on V582~Aur, and         
composed of the Digitized Sky Survey~2 blue (blue), \textit{WISE\/}      
12\,$\mu$m (green), and \textit{WISE\/} 22\,$\mu$m (red) images. The     
image suggests that the cometary-shaped clouds at the centre are         
located on the south-eastern boundary of a large complex of clouds,      
including LDN~1516, and are apparently exposed to disruptive effects     
from the south-eastern direction. Two luminous members of Aur~OB\,1      
from \citeauthor{humphreys1978}' \citeyearpar{humphreys1978} list        
are found within the area of the image. HD\,35633 is a B0.5\,IV type     
star, located at an angular distance of 0.51~deg from V582~Aur,          
corresponding to an 11.8~pc projected separation at 1.32~kpc. HD\,35653  
is a B0.5\,V type star, at a projected distance of some 23~pc from       
V582~Aur. The ultraviolet radiation of these stars may ionize and        
compress the clouds. The spectacular \ion{H}{II} region Sh2-234 is a     
background object at 2.8~kpc from the Sun. Spectroscopic and	         
photometric data, available for the B1\,V type star HD~281147 (see       
Fig.~\ref{fig_map1}) in \textit{VizieR\/}, suggest that this star is     
also a background object.					         

\section{Summary}
\label{Sect_5}
We have investigated the star-forming environment of the less studied FUor 
V582~Aur. Based on H$\alpha$ emission detected in slitless spectra 
and infrared excesses revealed by UKIDSS, \textit{Spitzer\/}, and \textit{WISE\/}
archival data we identified 68 candidate low-mass young stars in the 
$11.5\arcmin\times11.5\arcmin$ area centred on 
V582~Aur, among them two bona fide low-mass protostars. 
An optical colour--magnitude diagram of 35 members of the 
group, compared with pre-main-sequence evolutionary models, 
suggests that they represent a new, active star-forming 
subsystem of the Auriga~OB\,1 association, located at a distance of 
1.32~kpc from the Sun. The narrow-band H$\alpha$ image of the 
region, available in the IPHAS archive, reveals a system 
of bright-rimmed, cometary clouds associated with the newly identified
young stars. Our results suggest that star formation in these clouds 
might have been triggered by the radiation field of a few hot members 
of Aur~OB\,1. The bolometric luminosity of V582~Aur, based on archival 
photometric data and on the adopted distance, is 
$146 \la L_\mathrm{bol}/L_{\sun} \la 324$.

\section*{Acknowledgements}
  This research is based on observations with the 2.2-m telescope of
  the University of Hawaii. We thank Colin Aspin and Mark Willman for
  their interest and support. We are grateful to \'Agnes K\'osp\'al for 
  careful reading of the manuscript. This paper makes use of data obtained 
  as part of the INT Photometric H$\alpha$ Survey of the Northern
  Galactic Plane (IPHAS, www.iphas.org) carried out at the Isaac
  Newton Telescope (INT). The INT is operated on the island of La
  Palma by the Isaac Newton Group in the Spanish Observatorio del
  Roque de los Muchachos of the Instituto de Astrofisica de Canarias.
  All IPHAS data are processed by the Cambridge Astronomical Survey
  Unit, at the Institute of Astronomy in Cambridge. The bandmerged DR2
  catalogue was assembled at the Centre for Astrophysics Research,
  University of Hertfordshire, supported by STFC grant ST/J001333/1.
  This work also makes use of observations made with the \textit{Spitzer Space
  Telescope\/}, which is operated by the Jet Propulsion Laboratory,
  California Institute of Technology under a contract with NASA. This
  research was supported by ESTEC Contract No. 4000106398/12/NL/KML.
  This work was supported by the Hungarian Research Fund OTKA grants
  K81966 and K101393. This research has made use of the VizieR
  catalogue access tool, CDS, Strasbourg, France. The original
  description of the VizieR service was published in A\&AS 143,
  23. 

\begin{landscape}
\begin{table}
\caption{UKIDSS sources classified as candidate YSOs}
\label{table2}
\begin{center}
\begin{tabular}{cccccccl}
\hline
\hline
\noalign{\smallskip}
%\footnotesize{
   UGPS               &   J               &      H             &   K$_{\mathrm{s}}$   &    [3.6]  &     [4.5]  &   SST\,GLMC   & Note \\
\noalign{\smallskip}
\hline
\noalign{\smallskip}
 J052524.85+345800.3  &  17.590$\pm$0.021  &  16.529$\pm$0.015  &  15.720$\pm$0.020   &   14.826$\pm$0.056  &	14.281$\pm$0.046  &  G172.6121-00.3464 & \\ 
 J052525.94+345631.7  &  17.306$\pm$0.016  &  16.474$\pm$0.014  &  15.875$\pm$0.023   &   15.166$\pm$0.065  &	14.983$\pm$0.063  &  G172.6346-00.3572 & \\ 
 ~J052526.60+345756.5$^\mathrm{b}$  &  15.258$\pm$0.003  &  14.078$\pm$0.002  &  13.127$\pm$0.002   &   11.965$\pm$0.032  &	11.494$\pm$0.029  &  G172.6164-00.3421 & \\ 
 J052528.89+345627.6  &  16.367$\pm$0.008  &  15.047$\pm$0.004  &  13.967$\pm$0.004   &   13.031$\pm$0.048  &   12.465$\pm$0.036  &  G172.6412-00.3494 & \\
 J052529.76+345728.1  &  16.181$\pm$0.007  &  15.414$\pm$0.006  &  14.939$\pm$0.010   &   14.542$\pm$0.049  &	14.290$\pm$0.052  &  G172.6290-00.3376 & \\ 
 J052530.87+345744.5  &  15.992$\pm$0.006  &  15.198$\pm$0.005  &  14.656$\pm$0.008   &   14.065$\pm$0.054  &	13.884$\pm$0.041  &  G172.6273-00.3319 & \\ 
 J052532.03+345736.7  &  16.497$\pm$0.008  &  15.314$\pm$0.005  &  14.416$\pm$0.006   &   13.357$\pm$0.049  &	12.838$\pm$0.047  &  G172.6314-00.3298 & \\ 
 J052532.08+345721.8  &  17.016$\pm$0.013  &  16.269$\pm$0.012  &  15.723$\pm$0.020   &   14.891$\pm$0.060  &	14.426$\pm$0.053  &  G172.6349-00.3320 & \\ 
 J052532.08+345718.4  &  16.558$\pm$0.009  &  15.775$\pm$0.008  &  15.269$\pm$0.013   &   14.859$\pm$0.064  &	14.763$\pm$0.085  &  G172.6356-00.3325 & \\ 
 J052532.12+345734.7  &  17.197$\pm$0.015  &  16.223$\pm$0.012  &  15.549$\pm$0.017   &   15.007$\pm$0.104  &    $\cdots$         &  G172.6320-00.3299 &  \\ 
 J052532.51+345718.0  &  13.918$\pm$0.002  &  12.878$\pm$0.001  &  12.105$\pm$0.001   &   11.000$\pm$0.043  &	10.042$\pm$0.030  &  G172.6366-00.3313 & L1516-Ha\,3 \\ 
 J052532.60+345751.9  &  18.132$\pm$0.033  &  16.336$\pm$0.013  &  14.535$\pm$0.007   &   13.048$\pm$0.045  &   12.406$\pm$0.033  & G172.6289-00.3258 & \\
 J052532.73+345710.7  &  13.125$\pm$0.001  &  12.162$\pm$0.001  &  11.544$\pm$0.001   &   10.641$\pm$0.058  &	10.217$\pm$0.035  &  G172.6386-00.3319 & L1516-Ha\,4 \\ 
 ~J052532.74+345742.6$^\mathrm{b}$  &  16.038$\pm$0.006  &  14.609$\pm$0.003  &  13.516$\pm$0.003   &   12.154$\pm$0.047  & 11.600$\pm$0.031  &  G172.6314-00.3269 & \\ 
 J052533.03+345648.2  &  15.858$\pm$0.005  &  14.740$\pm$0.003  &  13.974$\pm$0.005   &   12.636$\pm$0.044  &	12.072$\pm$0.035  &  G172.6444-00.3345 & \\ 
 J052533.85+345752.1  &  16.598$\pm$0.009  &  15.745$\pm$0.008  &  15.144$\pm$0.012   &   14.266$\pm$0.053  &	13.806$\pm$0.042  &  G172.6313-00.3222 & \\
 J052534.08+345746.6  &  15.062$\pm$0.003  &  14.122$\pm$0.002  &  13.561$\pm$0.003   &   12.894$\pm$0.036  &	12.405$\pm$0.039  &  G172.6330-00.3225 & L1516-Ha\,6 \\ 
 J052536.40+345631.1  &  18.169$\pm$0.035  &  16.926$\pm$0.021  &  16.041$\pm$0.026   &   15.328$\pm$0.068  &	14.899$\pm$0.067  &  G172.6548-00.3277 & \\ 
 J052537.27+345734.7  &  16.165$\pm$0.007  &  15.406$\pm$0.006  &  14.862$\pm$0.009   &   13.980$\pm$0.039  &	13.521$\pm$0.043  &  G172.6418-00.3153 & \\ 
 J052538.17+345635.5  &  17.281$\pm$0.016  &  16.331$\pm$0.013  &  15.534$\pm$0.017   &   14.469$\pm$0.050  &	13.861$\pm$0.050  &  G172.6572-00.3220 & \\ 
 J052538.69+345635.1  &  16.154$\pm$0.007  &  15.394$\pm$0.006  &  14.825$\pm$0.009   &   14.108$\pm$0.042  &	13.808$\pm$0.045  &  G172.6582-00.3205 & \\ 
 J052542.12+345309.8  &  16.718$\pm$0.010  &  15.092$\pm$0.004  &  14.077$\pm$0.005   &   13.006$\pm$0.047  &	12.462$\pm$0.029  &  G172.7121-00.3427 & \\ 
 J052543.22+345243.1  &  18.313$\pm$0.039  &  17.237$\pm$0.029  &  16.564$\pm$0.043   &   16.326$\pm$0.120  &   16.018$\pm$0.108  &  G172.7203-00.3438 &  \\ 
 J052544.59+345357.7  &  17.715$\pm$0.023  &  16.920$\pm$0.022  &  16.378$\pm$0.037   &   15.987$\pm$0.088  &	15.840$\pm$0.087  &  G172.7057-00.3284 & \\ 
 J052545.12+345253.5  &  18.905$\pm$0.066  &  15.979$\pm$0.009  &  14.205$\pm$0.006   &   12.291$\pm$0.032  &   11.541$\pm$0.032  &  G172.7216-00.3368 & \\
 J052547.46+345228.3  &  19.162$\pm$0.083  &  16.984$\pm$0.023  &  14.953$\pm$0.010   &   12.970$\pm$0.054  &   12.323$\pm$0.039  &  G172.7318-00.3341 & \\
 J052548.08+345538.3  &  16.595$\pm$0.009  &  15.495$\pm$0.006  &  14.760$\pm$0.009   &   13.802$\pm$0.045  &	13.498$\pm$0.043  &  G172.6893-00.3028 & \\   
 J052548.45+345101.4  &  16.553$\pm$0.009  &  15.447$\pm$0.006  &  14.760$\pm$0.009   &   14.215$\pm$0.100  &   14.096$\pm$0.069  &  G172.7537-00.3448 & \\  
 J052550.09+345504.9  &  17.493$\pm$0.019  &  16.000$\pm$0.010  &  14.963$\pm$0.011   &   14.153$\pm$0.058  &	13.919$\pm$0.062  &  G172.7008-00.3023 & \\  	      
 J052551.33+345226.6$^\mathrm{a}$  &  17.463$\pm$0.019  &  15.697$\pm$0.007  &  14.410$\pm$0.007   & $\cdots$ &  $\cdots$  &  $\cdots$ \\
 ~~J052552.02+345808.2$^\mathrm{a,b}$  &      14.901$\pm$0.003 &  13.766$\pm$0.002  &  12.685$\pm$0.002   &   11.218$\pm$0.032  &	10.719$\pm$0.024  &  G172.6624-00.2683 & L1516-Ha\,8 \\ 	      
 J052553.87+345201.5  &  16.995$\pm$0.013  &  15.909$\pm$0.009  &  15.093$\pm$0.012   &   14.300$\pm$0.070  &	13.795$\pm$0.053  &  G172.7413-00.3141 & \\  	      
 J052554.39+345439.2  &  13.628$\pm$0.001  &  12.496$\pm$0.001  &  11.558$\pm$0.001   &   10.500$\pm$0.027  &	10.006$\pm$0.030  &  G172.7150-00.2941 & L1516-Ha\,9 \\ 	  
 J052554.52+345206.2  &  16.215$\pm$0.007  &  15.403$\pm$0.006  &  14.917$\pm$0.010   &   14.405$\pm$0.073  &	14.294$\pm$0.085  &  G172.7504-00.3175 & \\ 
 J052555.20+345201.2  &  17.561$\pm$0.020  &  16.712$\pm$0.018  &  16.114$\pm$0.029   &   15.266$\pm$0.083  &	14.936$\pm$0.073  &  G172.7529-00.3164 &  \\ 	      
 J052556.88+345144.2  &  13.530$\pm$0.001  &  12.649$\pm$0.001  &  11.899$\pm$0.001   &   10.973$\pm$0.038  &	10.438$\pm$0.027  & ~G172.7600-00.3143$^\mathrm{c}$  & L1516-Ha\,11  \\  
 J052556.97+345018.9  &  16.619$\pm$0.009  &  15.775$\pm$0.008  &  15.169$\pm$0.013   &   14.149$\pm$0.046  &	13.745$\pm$0.042  &  G172.7798-00.3273 & \\  	      
 J052600.97+345143.2  &  16.386$\pm$0.008  &  15.477$\pm$0.006  &  14.907$\pm$0.010   &   14.229$\pm$0.057  &	13.905$\pm$0.047  &  G172.7681-00.3029 &  \\ 	      
 J052602.74+345109.3  &  16.000$\pm$0.006  &  15.014$\pm$0.004  &  14.377$\pm$0.007   &   13.386$\pm$0.039  &	12.901$\pm$0.037  &  G172.7792-00.3031 & L1516-Ha\,12 \\ 	      
 J052608.38+345006.3  &  17.028$\pm$0.014  &  16.277$\pm$0.013  &  15.754$\pm$0.022   &   15.125$\pm$0.054  &	14.860$\pm$0.062  &  G172.8046-00.2969 & \\ 
 J052613.27+344936.1  &  14.817$\pm$0.003  &  13.940$\pm$0.002  &  13.396$\pm$0.003   &   12.501$\pm$0.047  &	12.137$\pm$0.038  &  G172.8209-00.2877 & \\ 
 ~J052613.58+345316.0$^\mathrm{a}$  &  17.950$\pm$0.030 & 16.698$\pm$0.019 & 15.664$\pm$0.021 &  $\cdots$ &  14.249$\pm$0.108&  G172.7708-00.2527  & \\ 
 ~J052613.92+345329.4$^\mathrm{a}$  &  16.387$\pm$0.008  &  14.786$\pm$0.004  &  13.660$\pm$0.004   &   12.865$\pm$0.043  &	12.401$\pm$0.037  &  G172.7684-00.2496 & \\ 
 J052614.68+345005.2  &  15.802$\pm$0.005  &  14.870$\pm$0.004  &  14.320$\pm$0.006   &   15.125$\pm$0.054  &	14.860$\pm$0.062  &  G172.8046-00.2969 & L1516-Ha\,14  \\ 	      
~ J052616.07+345320.0$^\mathrm{a}$  &  15.719$\pm$0.005  &  14.662$\pm$0.003  &  13.608$\pm$0.004   &   12.169$\pm$0.050  &	11.570$\pm$0.035  &  ~G172.7747-00.2449$^{c}$ & \\  	    
 ~J052616.37+345333.2$^\mathrm{a}$  &  13.410$\pm$0.001  &  11.948$\pm$0.001  &  10.781$\pm$0.001   &    9.053$\pm$0.080  &	 8.284$\pm$0.042  &  G172.7722-00.2421 & L1516-Ha\,15 \\ 
 J052616.80+345422.3  &  15.500$\pm$0.004  &  14.657$\pm$0.003  &  14.031$\pm$0.005   &   13.247$\pm$0.045  &   12.784$\pm$0.030  &  G172.7617-00.2332 &  L1516-Ha\,16 \\ 	      
 J052619.57+344929.5  &  16.103$\pm$0.007  &  15.180$\pm$0.005  &  14.621$\pm$0.008   &    $\cdots$            &   15.797$\pm$0.089  &  G172.7826-00.2363 &  \\  
\hline
\end{tabular}
\end{center}
\smallskip
\flushleft{\small
$^\mathrm{a}$Classified as a candidate galaxy in UKIDSS. $^\mathrm{b}$Classified as candidate YSO based on 
\textit{AllWISE\/} data \citep{Marton2016}. $^\mathrm{c}$SST\,GLMA identifier.}
\end{table}
\end{landscape}

\begin{landscape}
\begin{table}
\caption{Spitzer sources classified as candidate YSOs}
\label{table3}
\begin{center}
\begin{tabular}{cccccccl}
\hline
\hline
\noalign{\smallskip}
%\footnotesize{
 UGPS                    &   J               &      H             &   K$_\mathrm{s}$   &    [3.6]            &     [4.5]          & SST\,GLMC & Note \\
\noalign{\smallskip}
\hline
\noalign{\smallskip}
 J052527.87+345741.9 &  14.983$\pm$0.003 & 14.150$\pm$0.002 & 13.679$\pm$0.004  &  13.158$\pm$0.045  &  12.914$\pm$0.038 & G172.6222-00.3407 & L1516-Ha\,1 \\ 
 J052529.63+345755.1 &  14.591$\pm$0.002 & 13.700$\pm$0.002 & 13.181$\pm$0.002  &  12.729$\pm$0.034  &  12.464$\pm$0.030 & G172.6225-00.3337 & L1516-Ha\,2 \\
 J052529.74+345655.8 &  15.004$\pm$0.003 & 14.117$\pm$0.002 & 13.719$\pm$0.004  &  13.242$\pm$0.033  &  12.997$\pm$0.043 & G172.6364-00.3426 & \\
 J052534.55+345800.3 &  16.121$\pm$0.006 & 15.373$\pm$0.006 & 14.929$\pm$0.010  &  14.328$\pm$0.057  &  14.036$\pm$0.049 & G172.6308-00.3190 & \\
 J052534.98+345701.1 &  16.505$\pm$0.009 & 15.641$\pm$0.007 & 15.301$\pm$0.014  &  14.874$\pm$0.070  &  14.566$\pm$0.065 & G172.6452-00.3270 & \\
 J052543.31+345307.2 &  18.492$\pm$0.045 & 16.860$\pm$0.021 & 15.983$\pm$0.026  &  14.754$\pm$0.060  &  14.364$\pm$0.057 & G172.7150-00.3398 & \\
 J052544.56+345017.8 &  14.758$\pm$0.003 & 13.827$\pm$0.002 & 13.439$\pm$0.003  &  12.919$\pm$0.040  &  12.336$\pm$0.031 & G172.7563-00.3626 & L1516-Ha\,7 \\ 
 J052546.99+345244.3 &    $\cdots$	 & 17.018$\pm$0.024 & 14.569$\pm$0.008  &  12.425$\pm$0.048  &  11.374$\pm$0.040 & G172.7272-00.3329 & \\
 ~J052547.76+344950.9$^\mathrm{a}$  &     $\cdots$	 & 17.096$\pm$0.025 & 14.320$\pm$0.006  &  11.993$\pm$0.065  &  10.761$\pm$0.037 & G172.7687-00.3577 & \\
 J052555.37+345441.9 &  15.465$\pm$0.004 & 14.447$\pm$0.003 & 13.869$\pm$0.004  &  13.030$\pm$0.038  &  12.673$\pm$0.036 & G172.7162-00.2909 & L1516-Ha\,10 \\ 
 J052600.91+345413.1 &  15.949$\pm$0.006 & 14.974$\pm$0.004 & 14.549$\pm$0.008  &  13.633$\pm$0.044  &  13.173$\pm$0.038 & G172.7334-00.2797 & \\
 J052601.71+344933.5 &  16.706$\pm$0.010 & 15.898$\pm$0.009 & 15.473$\pm$0.017  &  14.961$\pm$0.055  &  14.604$\pm$0.050 & G172.7993-00.3209 & \\
 J052607.10+345425.9 &  14.791$\pm$0.003 & 13.959$\pm$0.002 & 13.505$\pm$0.003  &  12.916$\pm$0.045  &  12.544$\pm$0.036 & G172.7423-00.2601 & L1516-Ha\,13 \\ 
 J052611.65+344904.5 &  16.640$\pm$0.010 & 15.909$\pm$0.009 & 15.479$\pm$0.018  &  14.580$\pm$0.056  &  14.202$\pm$0.041 & G172.8250-00.2972 & \\
 J052618.28+344716.9 &  15.480$\pm$0.004 & 14.666$\pm$0.003 & 14.198$\pm$0.006  &  13.557$\pm$0.038  &  13.130$\pm$0.033 & G172.8625-00.2951 & \\
 J052619.46+344943.9 &  15.410$\pm$0.004 & 14.501$\pm$0.003 & 14.045$\pm$0.005  &  13.392$\pm$0.048  &  13.079$\pm$0.038 & G172.8309-00.2689 & \\
 J052619.67+344923.7 &  16.267$\pm$0.007 & 15.405$\pm$0.006 & 14.956$\pm$0.011  &  14.465$\pm$0.065  &  14.035$\pm$0.041 & G172.8359-00.2715 & \\
\hline
\end{tabular}
\end{center}
\smallskip
\flushleft{\small
$^\mathrm{a}$Classified as a candidate galaxy in UKIDSS.}
\end{table}
\end{landscape}

\bibliographystyle{mnras}
\bibliography{v582}

\bsp	% typesetting comment
\label{lastpage}

\end{document}